\documentclass[a4paper, 10pt]{article}

\usepackage[utf8]{inputenc}
\usepackage[T1]{fontenc}
\usepackage{CJK}

\usepackage{amsmath}
\usepackage{amsthm}
\usepackage{amsfonts}
\usepackage{amssymb}

\usepackage{xspace}
\usepackage[english]{babel}
\usepackage{bigfoot}
\usepackage{textcomp}
\usepackage[normalem]{ulem}
\usepackage{framed}
\usepackage{authblk}
\usepackage{placeins}

\usepackage{xcolor}

\def\mytitle{Wrangling Messy CSV Files by Detecting Row and Type Patterns}

\providecommand{\keywords}[1]{\\\vskip.25\baselineskip\noindent\textbf{\textit{Keywords 
			---}} #1}

\usepackage{natbib}
\usepackage{hyperref}
\hypersetup{
    unicode=false,          
    pdftoolbar=true,        
    pdfmenubar=true,        
    pdffitwindow=false,     
    pdfstartview={FitH},    
    pdftitle={\mytitle},    
    pdfauthor={G.J.J. van den Burg and A. Nazabal and C. Sutton}, 
    pdfcreator={G.J.J. van den Burg},   
    pdfproducer={G.J.J. van den Burg}, 
    pdfkeywords={Data Wrangling, Data Parsing, Comma Separated Values}, 
    pdfnewwindow=true,      
    colorlinks=true,       
    linkcolor=black,          
    citecolor=black,        
    filecolor=magenta,      
    urlcolor=cyan           
}

\usepackage{graphicx}
\graphicspath{{./images/}}

\usepackage{algorithm}
\usepackage{algpseudocode}

\usepackage[%
hang,
margin=10pt,
font=small,
labelfont=bf,
labelsep=period]{caption}
\usepackage{subcaption}

\usepackage[a4paper]{geometry}

\newcommand{\bft}[1]{\mathbf{#1}}
\newcommand{\bfs}[1]{\boldsymbol{#1}}

\newcommand{\abs}[1]{\left|#1\right|}

\DeclareMathOperator*{\argmax}{arg\,max}

\newtheorem{theorem}{Theorem}[section]

\newtheorem{definition}[theorem]{Definition}

\bibpunct{(}{)}{;}{a}{,}{,}

\title{\mytitle}
\author[1]{Gerrit J.J. van den Burg}
\author[1]{Alfredo Naz{\'a}bal}
\author[1,2,3]{Charles Sutton}
\affil[1]{The Alan Turing Institute, London, UK}
\affil[2]{Google, Inc. Mountain View, CA, USA}
\affil[3]{School of Informatics, The University of Edinburgh, UK}
\date{\today}

\def\mycitep{\citep}
\def\mycitet{\citet}

\begin{document}

\maketitle

\begin{abstract}
	It is well known that data scientists spend the majority of their time 
	on preparing data for analysis. One of the first steps in this 
	preparation phase is to load the data from the raw storage format.  
	Comma-separated value (CSV) files are a popular format for tabular 
	data due to their simplicity and ostensible ease of use. However,   
	formatting standards for CSV files are not followed consistently,
	so each file requires manual inspection and potentially repair 
	before the data can be loaded, an enormous waste of human effort for a 
	task that should be one of the simplest parts of data science. The 
	first and most essential step in retrieving data from CSV files is 
	deciding on the dialect of the file, such as the cell delimiter and
	quote character. Existing dialect detection approaches are few and 
	non-robust.  In this paper, we propose a dialect detection method 
	based on a novel measure of data consistency of parsed data files.
	Our method achieves %
	97\%
 overall accuracy 
	on a large corpus of real-world CSV files and improves the accuracy on 
	messy CSV files by almost %
	22\%

	compared to existing approaches, including those in the Python 
	standard library.
\end{abstract}

\clearpage

\noindent\emph{CSV is a textbook example of how \emph{not} to design a textual 
	file format.}%
\begin{flushright}
--- The Art of Unix Programming, \mycitet{raymond2003art}.%
\end{flushright}%

\section{Introduction}

The goal of data science is to extract valuable knowledge from data through 
the use of machine learning and statistical analysis. Increasingly however, it 
has become clear that in reality data scientists spent the majority of their 
time importing, organizing, cleaning, and wrangling their data in preparation 
for the analysis 
\mycitep{dasu2003exploratory,lohr2014for,kandel2011wrangler,crowdflower2016data,kaggle2017state}.  
Collectively this represents an enormous amount of time, money, and talent. As 
the role of data science is expected to only increase in the future, it is important 
that the mundane tasks of data wrangling are automated as much as possible.  
It has been suggested that one reason that data scientists spent the majority 
of their time on data wrangling issues is due to what could be called the 
\emph{double Anna Karenina principle}: ``every messy dataset is messy in its 
own way, and every clean dataset is also clean in its own way'' 
\mycitep{sutton2018data}.\footnote{This problem is related to the principle of 
	the fragility of good things \mycitep{arnold2003catastrophe}.} Because 
of the wide variety of data quality issues and data formats that exist 
(``messy in its own way''), it is difficult to re-use data wrangling scripts 
and tools, perhaps explaining the manual effort required in data wrangling.


This problem can be observed even in the earliest and what might be considered  
the simplest stages of the data wrangling process, that of loading and parsing 
the data in the first place.  In this work, we focus as an example on 
comma-separated value (CSV) files, which despite their deceptively simple 
nature, pose a rich source of formatting variability that frustrates data 
parsing. CSV files are ubiquitous as a format for sharing tabular data on the 
web; based on our data collection, we conservatively estimate that GitHub.com 
alone contains over 19 million CSV files.  Open government data repositories 
make increasingly more datasets available and often present their data in CSV 
format.\footnote{\mycitet{mitlohner2016characteristics} survey 200,000 CSV 
	files from open government data portals.} Advantages of CSV files 
include their simplicity and portability, but despite some standardization 
effort \citep[RFC 4180;][]{shafranovich2005common}, a wide variety of subtly 
incompatible variations of CSV files exist, including the 39 different 
dialects among the CSV files in our data. For example, we observe that values 
can be separated by commas, semicolons, spaces, tabs, or any other character, 
and can be surrounded by quotation marks or apostrophes to guard against 
delimiter collision or for no reason at all. Because so many variations of the 
CSV file format exist, manual inspection is often required before a file can 
be loaded into an analysis program.
 
\begin{figure}[tb]
	\centering
	\includegraphics[height=8.9cm]{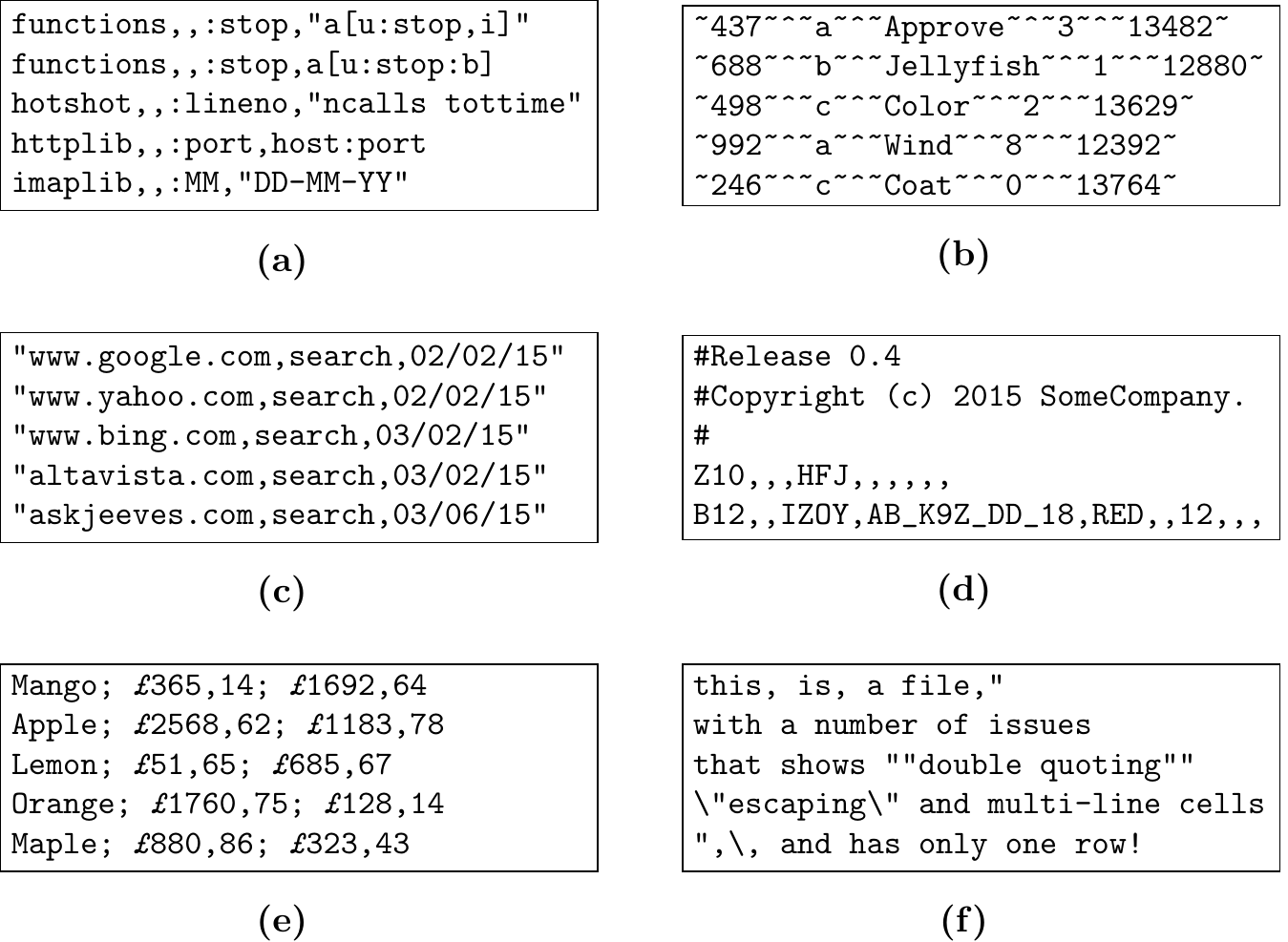}
	\caption{Illustration of some of the variations of real-world CSV 
		files. See the main text for a description of each of the 
		files. \label{fig:csv_weird}}
\end{figure}

Every messy CSV file is indeed messy in its own way, as we see from  analysing
a corpus of over 18,000 files in our study (Section~\ref{sec:comparison}).
Figure~\ref{fig:csv_weird} illustrates a few of the variations and problems 
that real-world CSV files exhibit by presenting simplifications of real-world 
files encountered during this study.  Figure~\ref{fig:csv_weird}(a) 
illustrates a normal CSV file that uses comma as the delimiter and has both 
empty and quoted cells.  Figure~\ref{fig:csv_weird}(b) shows a variation 
encountered in this study that uses the caret symbol as delimiter and the 
tilde character as quotation mark.  Next, Figure~\ref{fig:csv_weird}(c) 
illustrates an \emph{ambiguous} CSV file: the entire rows are surrounded with 
quotation marks, implying that the correct interpretation is a single column 
of strings.  However, if the quotation marks are stripped a table appears 
where values are separated by the comma.  Figure~\ref{fig:csv_weird}(d) 
illustrates a file with comment lines that have the hash symbol as a prefix.  
Figure~\ref{fig:csv_weird}(e) is adapted from \mycitet{dohmen2017multi} and 
illustrates a file where multiple choices for the delimiter result in the same 
number of rows and columns (the semicolon, space, comma, and pound sign all 
yield three columns).  Finally, Figure~\ref{fig:csv_weird}(f) illustrates a 
number of issues simultaneously: quote escaping both by using an escape 
character and by using double quotes, delimiter escaping, and multi-line 
cells.  

In this paper, we present a method for automatically determining the formatting
parameters, which we call the \emph{dialect}, of a CSV file. Our method is based
on a novel consistency measure for parsed data files that allows us to search 
the space of dialects for one under which the parsed data is most  consistent. 
By \emph{consistency} here we consider primarily (a) the shape of the parsed 
data, which we capture using an abstraction called \emph{row patterns}, and 
(b) the data types of the cells, such as integers or strings.  This aims to 
capture how a human analyst might identify the dialect: searching for a 
character that results in regular row patterns and using knowledge of what 
real data ``looks like''. 

It may surprise the reader that CSV parsing is an open problem. However, even 
though almost every programming language provides functionality to read data 
from CSV files, very few are robust against the many issues encountered in 
real-world CSV files. In nearly every language the user is responsible for 
correctly setting the parameters of the parsing procedure by manually 
providing function arguments such as the cell delimiter, the quote character, 
whether or not headers exist, how many lines to skip before the data table 
starts, etc.  To the best of our knowledge, Python is the only programming 
language whose standard library supports automatically detecting some of these 
parsing parameters through a so-called dialect ``sniffer''.  In all other 
languages the user is responsible for correctly choosing the parameters when 
the CSV file doesn't adhere exactly to the format in the RFC 4180 
specification, or otherwise risks that the file is loaded incorrectly or not 
at all. Moreover, we shall see below that even the Python dialect sniffer is 
not robust against many real-world CSV files.  This means that in practice 
almost every file requires manual inspection before the data can be loaded, 
since it may contain a non-standard feature.  Automatic detection of the 
dialect of CSV files aims to alleviate this issue.

In the formalism for CSV parsing we present in this paper, we distinguish two 
aspects of what makes the format of a file ``messy''. First, CSV files have a 
dialect that captures how tabular data has been converted to a text file. This 
dialect contains the delimiter, the quote character, and the escape character.  
Knowing this dialect is the first step to importing the data from the file as 
it provides all the information necessary to convert the raw text file into a 
matrix of strings.  Second, CSV files may contain headers, have additional 
comment text (potentially indicated by a prefix character), contain multiple 
tables, encode numeric data as strings, or contain any of the myriad other 
issues that affect how the tabular data is represented. Thus while the dialect 
is sufficient to recover a matrix of strings from the file, the full CSV 
parsing problem contains additional steps to recover the original data.

The problem of dialect detection, which is our focus here, is a subproblem of 
CSV parsing in general. The broader CSV parsing problem includes file encoding 
detection, dialect detection, table detection, header detection, and cell type 
detection.  We restrict ourselves to dialect detection for two reasons. First, 
while encoding detection for CSV files may be slightly different than the 
general encoding detection problem due to different character frequencies, we 
consider this problem sufficiently solved for our purposes and employ the  
method of \mycitet{li2001composite}.  With this in mind, the next problem to 
solve is that of dialect detection.  Second, the vast majority of CSV files 
have only a single table \mycitep{mitlohner2016characteristics} and thus 
solving dialect detection is a more pressing issue than table detection.

A CSV file is created by some CSV ``formatter'' (either a program or a human) 
using a particular dialect. To load the file and decide on the dialect a human 
analyst would use their understanding of the CSV format and its many 
variations, an understanding of what the data embedded in the file represents, 
and potentially years of experience of dealing with messy CSV files.  
Automating this process is non-trivial: we receive a text file from an unknown 
source, created with an unknown formatter using an unknown dialect, that 
contains unknown data, and are asked to choose the parameters that allow a 
faithful reconstruction of the original data.  Considering the large number of 
unknowns in this process, it may not be  surprising that many programming 
languages do not attempt to automate this process at all and instead leave the 
user to deal with the issue. It may also explain why only very little 
attention has been paid to this problem in the literature.

This paper is structured as follows. In Section~\ref{sec:related_work} we 
present an overview of related work on both table detection and CSV parsing.  
Next, Section~\ref{sec:problem_statement} gives a formal mathematical 
description of CSV parsing that presents dialect detection as an inverse 
problem.  Our proposed data consistency measure for dialect detection is 
presented in Section~\ref{sec:consistency}.  Results of a thorough comparison 
of our method with the few existing alternatives are presented in 
Section~\ref{sec:comparison}.  Section~\ref{sec:conclusion} concludes the 
paper.

\section{Related Work}%
\label{sec:related_work}

Only very few publications have paid any attention to the problem of CSV 
parsing.  \mycitet{mitlohner2016characteristics} explore a large collection 
(200K) of CSV files extracted from open data platforms of various governments.  
The authors use a number of heuristics to explore and parse the files. Dialect 
detection is done with the Python built-in CSV sniffer mentioned above, but 
the accuracy of this detection is not evaluated.  Despite the extensive 
analysis of this large corpus of CSV files, the authors do not present a novel 
CSV parser or dialect detection method based on their heuristics. 

A recent paper that does present a novel CSV parser and specifically addresses 
the problem of messy CSV files is that of \mycitet{dohmen2017multi}. They 
present HypoParsr, a so-called ``multi-hypothesis'' CSV parser that builds a 
tree of all possible parser configurations and subsequently scores each 
configuration with a heuristic method. The authors evaluate their parser on 64 
files with known ground truth and a corpus of 15,000 files from the UK open 
government portal without ground truth. While they achieve a high parsing 
``accuracy'' on the latter corpus of CSV files, this does not necessarily  
reflect a correct parsing of the CSV files due to the absence of ground truth.  
In our preliminary analysis of this parser we observed that its implementation 
is rather fragile and does not return a parsing result for many messy CSV 
files.  Moreover, on several files from the UK open government portal we 
observed that while the parser returned without error, it did in fact quietly 
drop several rows from the result. This illustrates the difficulty of the CSV 
parsing problem: without a large enough set of real-world messy CSV files it 
is hard to create a parser that is robust against all the problems and 
variations that can occur.

A closely related subject to CSV parsing and dialect detection is that of 
detecting tables in free text. Early work on this topic includes the work of 
\mycitet{ng1999learning} who address the problem of identifying the 
\emph{location} of a table in free text. They use a decision tree classifier 
based on features extracted from the content of the text, such as whether or 
not a character is next to whitespace, or a character is a symbol, etc. Later 
work by \mycitet{pinto2003table} applies a similar strategy for constructing 
features but instead uses conditional random fields and expands the problem by 
also identifying the semantic role of each row in the identified table (i.e.  
header, data row, etc.).  However, CSV parsing differs from identifying tables 
in free text because CSV files have more structure and flexibility than free 
text tables. CSV files use a specific character to delimit cells and can 
employ the quoting mechanism to let cells span multiple lines and guard 
against delimiter collision. This explains why the methods mentioned above for 
detecting tables in free text cannot be readily applied to CSV parsing.

Other work related to CSV parsing is the work on DeExcelerator by 
\cite{eberius2013deexcelerator}. This work focuses mainly on spreadsheets in 
Excel format, but can also handle CSV files when the correct dialect is 
provided by the user. The DeExcelerator program then extracts tables from the 
spreadsheet and in the process performs header recognition, data type 
recognition, and value extrapolation, and other operations based on heuristic 
rules. In follow-up work, \mycitet{koci2016machine} present a method for 
classifying the role of each cell in the spreadsheet (i.e. attribute, data, 
header, etc.) using surface-level features derived from the formatting of the 
text and classification methods such as decision trees and support vector 
machines.  While the DeExcelerator package offers no method for detecting the 
dialect of CSV files, the methods for table and header detection are suitable 
to the general CSV parsing problem outlined above.

More broadly, there is some work on data wrangling in particular that can be 
considered here. \mycitet{fisher2008dirt} present the PADS system for 
retrieving structured data from ad-hoc data sources such as log files of 
various systems. However, the authors explicitly mention that CSV files are 
not a source of ad-hoc data. Well-known work on the more general data 
wrangling problem is that of \mycitet{kandel2011wrangler} on the Wrangler 
system. Here the human-machine interaction of data wrangling is considered and 
a system is proposed to assist the human analyst. In follow-up work, 
\mycitet{guo2011proactive} provide a method for automatically suggesting data 
transformations to the analyst and additionally provide a ``table suitability 
metric'' that quantifies how well a table corresponds to a relational format.  
The relational format is also known as attribute-value format or ``tidy data'' 
\mycitep{wickham2014tidy} and it is often the goal of the data wrangling 
process. In our experiments we compare our method to the table suitability metric
of \mycitet{guo2011proactive}.

Finally, it is worth mentioning the few efforts that have aimed to solve the 
problem of CSV parsing at the source by proposing extensions or variations on 
the format that address some of the known issues with CSV files. A study on 
the use and future of CSV files on the web was performed by a working group  
of the World Wide Web Consortium \mycitep{tennison2016csv}. One of the results 
of this working group is a proposal to provide metadata about a CSV file 
through an accompanying JSON \mycitep{crockford2006application} description 
file \mycitep{tennison2015metadata}. A similar proposal is the CSV Dialect 
specification by \mycitet{frictionless2017csv} that recommends storing the 
dialect of the CSV file in a separate JSON object.  While these 
recommendations could certainly address some of the issues of CSV parsing, it 
requires users to specify a second file and maintaining it next to the CSV file 
itself. Moreover, it does not address the issues of the existing messy CSV 
files. Alternatives such as the CSVY format \mycitep{rovegno2015csvy} propose 
to add a YAML \mycitep{evans2001yaml} header with metadata. While this does 
combine the metadata and tabular data in a single file, it requires special 
import and export tools that may limit the adoption of these formats.

\section{Problem Statement}%
\label{sec:problem_statement}

We present a formal mathematical definition of CSV file parsing based on a 
generative process for how the file was constructed from the underlying data.  
This formal framework will define the language we subsequently use to present 
our methodology and will clarify the problem of CSV parsing in general. Within 
this framework dialect detection is presented as an inverse problem.  This is 
a natural framing of the problem since we are interested in identifying the 
unknown dialect that was used to produce an observed outcome. 

Let $\Sigma$ denote a finite alphabet of characters in some encoding $E$ and 
let $\Sigma^*$ be all the possible strings in this alphabet.\footnote{The set 
	$\Sigma^*$ is known as the Kleene closure 
	\mycitep{kleene1951representation}.} Then a CSV file is an element 
$\bft{x} \in \Sigma^*$. A CSV file contains $N$ tuples $\bft{t}_i$ that 
represent elements of a product set, thus
\begin{equation}
	\bft{t}_i \in \mathcal{V}_1 \times \mathcal{V}_2 \times 
	\cdot\cdot\cdot \times \mathcal{V}_{L_i}.
\end{equation}
where the sets $\mathcal{V}_j$ are the domains of values in the tuples 
\mycitep{codd1970relational}. These domains represent the sets that the values 
belong to, i.e. that floating point numbers have the domain $\mathcal{V} = 
\mathbb{R}$, for instance.  Note that the length of a tuple is given by $L_i$.  
Since CSV files can contain comments or multiple tables we cannot assume that 
the length of tuples is constant throughout the file. The collection of all 
tuples in the file is given by the array $\bft{T} = [\bft{t}_1, \ldots, 
\bft{t}_N]$.

Next we define the \emph{dialect} of CSV files. Conceptually, we can think of 
a CSV file as being generated by a \emph{formatter} that takes the original 
data and converts it to its CSV file representation.  The formatter has 
knowledge of the dialect $\bfs{\theta}$ that has been used to create the CSV 
file, and it consists of two stages. In Stage~1 the formatter converts the 
data $\bft{T}$ to a string-only representation $\bft{C}$ where the domains in 
the tuples are the string domain (i.e. $\Sigma^*$). In Stage~2 this 
string-only representation is converted to a CSV file $\bft{x}$.  When we wish 
to retrieve the data from the CSV file the information about both the 
formatter and the dialect has been lost.  This leads us to the following 
definition.

\begin{definition}[Dialect]
	\label{def:dialect}
	Given a CSV file $\bft{x} \in \Sigma^*$, the \emph{dialect} 
	$\bfs{\theta}$ represents all parameters needed for a one-to-one 
	mapping from an array of tuples of strings $\bft{C} = [\bft{c}_1, 
	\ldots, \bft{c}_N]$ to the file $\bft{x}$, thus
	\begin{equation}
		f_{\bfs{\theta}}(\bft{C}) = \bft{x}.
	\end{equation}
	Here $f_{\bfs{\theta}}$ represents Stage~2 of the formatter. The 
	inverse operation $f_{\bfs{\theta}}^{-1}(\bft{x}) = \bft{C}$ 
	corresponds to converting the file $\bft{x}$ to an array of tuples of 
	strings.
\end{definition}

We may further define a function $g$ that is used in Stage~1 of the formatter 
to convert the data to the string representation,
\begin{equation}
	g(\bft{T}) = \bft{C}.
\end{equation}
The inverse of this function, $g^{-1}(\bft{C})$, returns the elements in the 
tuples to their original representation and is in practice often achieved 
through type casting.  In practice one may additionally wish to specify an 
inverse operation $h^{-1}(\bft{T})$ that extracts from $\bft{T}$ the actual 
\emph{tabular data} by identifying the location of rectangular tables in 
$\bft{T}$, detecting the presence of column headers, removing unnecessary 
comments, etc. This leads us to the following general definition.

\begin{definition}[CSV Parsing]
	Let $\bft{x} \in \Sigma^*$ be a CSV file. The problem of \emph{CSV 
		parsing} then reduces to evaluating
	\begin{equation}
		h^{-1}(g^{-1}(f_{\bfs{\theta}}^{-1}(\bft{x}))).
	\end{equation}
\end{definition}

Thus we see that the first step in parsing CSV files is indeed the detection 
of the correct dialect $\bfs{\theta}$. In the remainder of this paper we will 
therefore focus our attention on this step. In the process we will often use 
the parsing function $f^{-1}_{\bfs{\theta}}$ to parse the file for a given 
dialect.  Although the exact definition of this function is not the main focus 
of the paper, it is worth mentioning that we base our implementation on the 
CSV parser in the Python standard library. Minor modifications of this code 
were made to simplify the dialect and handle certain edge cases differently, 
see Appendix~\ref{app:parser} for more details.

In this work we consider a dialect of three components: the delimiter 
($\theta_d$), the quote character ($\theta_q$), and the escape character 
($\theta_e$). The \emph{delimiter} is used to separate cells (i.e. values in 
the tuple), the \emph{quote character} is used to enclose cells that may 
contain the delimiter, a newline character, or neither, and the \emph{escape 
	character} can be used to achieve nested quotation marks. Each of 
these parameters can be absent in a CSV file, thus $\theta_d, \theta_q, 
\theta_e \in \Sigma \cup \{ \varepsilon \}$ with $\varepsilon$ the empty 
string (note that $\theta_d = \varepsilon$ for a file with a single column of 
data).\footnote{Thus we restrict ourselves to CSV files where these parameters 
	are all a single character. CSV files that use multi-character 
	delimiters do exist, but are extremely rare.} While some existing 
parsers include other components of the dialect, such as whether or not nested 
quotes are escaped with double quotes, we found that it was possible to 
formulate  $f^{-1}_{\bfs{\theta}}$ with only the three components given above.  
Moreover, some of these other components are more accurately described as 
parameters to the functions $g^{-1}$ or $h^{-1}$.

\section{A Consistency Measure for Dialect Detection}
\label{sec:consistency}

Having illustrated above the difficulty of dialect detection in general, we 
present here our solution based on two components: row length patterns and 
data type inference. The main idea that if a CSV file is parsed with the correct
dialect rather than an incorrect one, the resulting tuples in the parsed data
will appear more consistent. By \emph{consistency} here we consider primarily
two aspects: (a) the parsed rows should have similar length, and (b) cells 
within the tuples should have the same data type, such as integers or strings.
To that end, we propose a measure of consistency over parsed data, which has two 
components, one that measures each of these two kinds of consistency.  Then we 
search the space of possible dialects for the dialect in which the parsed file 
is most consistent. This aims to capture both aspects of how a human analyst 
would identify the dialect: searching for a character that results in regular 
row patterns and using knowledge of what real data ``looks like''. 

More formally, we associate with each dialect $\bfs{\theta}$ a consistency 
measure $Q(\bft{x}, \bfs{\theta}) = P(\bft{x}, \bfs{\theta}) \cdot T(\bft{x}, 
\bfs{\theta})$, where $P$ is a \emph{pattern score} and $T$ is a \emph{type 
	score}. The estimate of the correct dialect is then obtained as
\begin{equation}
	\label{eq:problem}
	\bfs{\hat{\theta}} = \argmax_{\bfs{\theta} \in \Theta_{\bft{x}}} 
	Q(\bft{x}, \bfs{\theta}).
\end{equation}
Algorithm~\ref{alg:heuristic} shows pseudocode for our search algorithm. The 
algorithm uses the fact that the type score $T(\bft{x}, \bfs{\theta})$ is 
between 0 and 1 to speed up the search. In specific instances multiple 
dialects can receive the same value for $Q(\bft{x}, \bfs{\theta})$ due to 
limitations in the type score.  However, some of these ties can be broken 
reliably and we will expand on this below as well.  

\begin{algorithm}[tb]
	\footnotesize
	\caption{Dialect Detection \label{alg:heuristic}}
	\begin{algorithmic}[1]
		\Function{DetectDialect}{$\bft{x}$}
		\State $\Theta_{\bft{x}} \gets$ \Call{GetDialects}{$\bft{x}$}
		\State $\mathcal{H} \gets \emptyset$
		\State $Q_{max} \gets -\infty$
		\For {$\bfs{\theta} \in \Theta_{\bft{x}}$}
			\State $P(\bft{x}, \bfs{\theta}) \gets$ 
			\Call{PatternScore}{$\bft{x}$, $\bfs{\theta}$}
			\If {$P(\bft{x}, \bfs{\theta}) < Q_{max}$}
				\State \textbf{continue}
			\EndIf
			\State $T(\bft{x}, \bfs{\theta}) \gets$	
			\Call{TypeScore}{$\bft{x}$, $\bfs{\theta}$}
			\State $Q(\bft{x}, \bfs{\theta}) \gets P(\bft{x}, 
			\bfs{\theta})T(\bft{x}, \bfs{\theta})$
			\If {$Q(\bft{x}, \bfs{\theta}) > Q_{max}$}
				\State $\mathcal{H} \gets \{ \bfs{\theta} \}$
				\State $Q_{max} \gets Q(\bft{x}, \bfs{\theta})$
			\ElsIf {$Q(\bft{x}, \bfs{\theta}) = Q_{max}$}
				\State $\mathcal{H} \gets \mathcal{H} \cup \{ 
				\bfs{\theta} \}$
			\EndIf
		\EndFor
		\If {$\abs{\mathcal{H}} = 1$}
			\State \Return $\bfs{\theta} \in \mathcal{H}$
		\Else
			\State \Return \Call{TieBreaking}{$\mathcal{H}$}
		\EndIf
		\EndFunction
	\end{algorithmic}
\end{algorithm}

\subsection{Pattern Score}

\begin{figure}[tb]
	\centering
	\includegraphics[width=\textwidth]{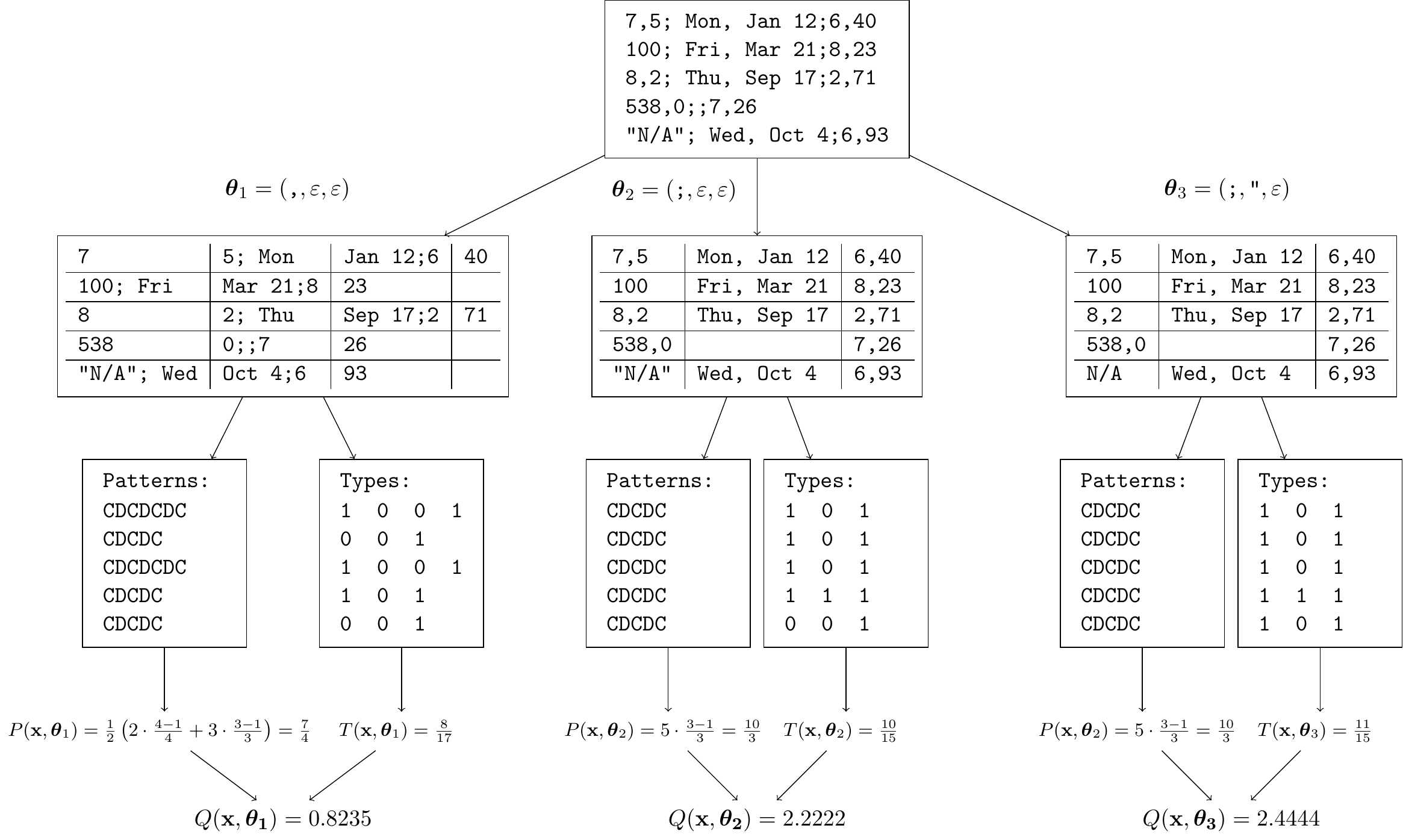}
	\caption{Illustration of the data consistency measure for different 
		dialect on a constructed example. The figure shows how 
		different dialects can result in different row patterns and 
		pattern scores, as well as different type scores. The 
		difference between $\bfs{\theta}_2$ and $\bfs{\theta}_3$ is 
		due to the fact that the string {\texttt N/A} belongs to a 
		known type, but the string {\texttt "N/A"} does not.}
	\label{fig:score_ill}
\end{figure}

The pattern score is the main driver of the data consistency measure. It is 
based on the observation that since CSV files generally contain tables, we 
expect to find rows with the same number of cells when we select the correct 
dialect. For a given dialect we therefore parse the file and determine the 
number of cells in each row. This parsing step takes nested quotes and escape 
characters into account, and returns for each row a so-called \emph{row 
	pattern}. See Figure~\ref{fig:score_ill} for an illustration. The row 
patterns capture the repeated pattern of cells and delimiters and interpret 
quotes and escape characters where needed while abstracting away the content 
of the cells.\footnote{This process interprets the quotes as defined by the 
	dialect, but it can occur that a spurious quote character remains in 
	the row pattern when an incorrect dialect is chosen. In this case, we 
	consider patterns such as \verb+CDCQCDC+ -- where \verb+Q+ denotes the 
	quote character, \verb+D+ the delimiter, and \verb+C+ any other 
	character -- to be distinct from \verb+CDCDC+.}  Notice how in 
Figure~\ref{fig:score_ill} the parsing result for $\bfs{\theta}_1$ gives 
different row patterns than for $\bfs{\theta}_2$ and $\bfs{\theta}_3$.  Each 
distinct row pattern $k$ has a length $L_k$, that is exactly one higher than 
the number of delimiters in the pattern. The number of times row pattern $k$ 
occurs in the file is called $N_k$ and the total number of distinct row 
patterns is denoted by $K$.

These properties are used to define the pattern score, as follows
\begin{equation}
	\label{eq:pattern_score}
	P(\bft{x}, \bfs{\theta}) = \frac{1}{K} \sum_{k=1}^K N_k \frac{L_k - 
		1}{L_k}.
\end{equation}
The pattern score is designed to favour row patterns that occur often and 
those that are long, and also to favour fewer row patterns. The reasoning 
behind this is that when the correct row pattern is chosen, we expect to 
observe few distinct row patterns, and that the ones that we do observe occur 
often (i.e.  in tables).  Figure~\ref{fig:score_ill} illustrates this aspect, 
as the correct delimiter yields a single row pattern that occurs throughout 
the entire file.  By including the row length ratio we favour longer row 
patterns over shorter ones. This is included because a long pattern indicates 
a regular pattern of delimiters and cells, whereas a short pattern might 
indicate an incorrectly chosen delimiter. 

While this function works well for CSV files that contain tables, it gives a 
value of $0$ when the entire file is a single column of data. To handle these 
files in practice, we replace the numerator by $\max\{\alpha, L_k - 1\}$ with 
$\alpha$ a small constant. The value of $\alpha$ must be chosen such that 
single-column CSV files are detected correctly, while avoiding false positive 
results that assume regular CSV files are a single column of messy data. It 
was found empirically that $\alpha = 10^{-3}$ achieves this goal well.

\subsection{Type Score}

While the pattern score is the main component of the data consistency measure, 
Figure~\ref{fig:score_ill} shows that the type score is essential to obtaining 
state-of-the-art results.  The goal of the type score is to act as a proxy for 
understanding what the cells of the file represent, capturing whether a 
dialect yields cells that ``look like real data''.  To do this, the type score 
measures the proportion of cells in the parsed file that can be matched to a 
data type from a set $\mathcal{T}$.

In our implementation, $\mathcal{T}$ includes empty cells, numbers in various 
formats, URLs, email addresses, percentages, currency values, times, dates, 
and alphanumeric strings (see Appendix~\ref{app:types}). We use regular 
expressions to detect these types, and denote the mapping from a string 
$\bft{z}$ to a type by $\mu(\bft{z})$. Then the type score is defined as
\begin{equation}
	\label{eq:type_score}
	T(\bft{x}, \bfs{\theta}) = \frac{1}{M} \sum_{i=1}^N \sum_{j=1}^{L_i} 
	I[\mu(c_{i,j}) \in \mathcal{T}]
\end{equation}
where $M$ is the total number of cells in the parsing result $\bft{C}$, 
$c_{i,j}$ is the $j$-th value of tuple $\bft{c}_i$, and $I[\cdot]$ denotes the 
indicator function that returns 1 if its argument is true and 0 otherwise.  
While the type detection algorithm can identify many data types, it is 
infeasible to encapsulate all possible data types. Thus, the situation may 
arise where no type can be detected for any cell in the parsing result. This 
would result in a consistency measure $Q(\bft{x}, \bfs{\theta}) = 0$, even 
though the pattern score may give high confidence for a dialect.  To avoid 
this problem in practice, we replace $T(\bft{x}, \bfs{\theta})$ by 
$\tilde{T}(\bft{x}, \bfs{\theta}) = \max\{\beta, T(\bft{x}, \bfs{\theta})\}$, 
with $\beta = 10^{-10}$.

\subsection{Tie breaking}

Unfortunately the value of $Q(\bft{x}, \bfs{\theta})$ can be the same for 
different dialects even in the presence of the type score, due to the fact 
that the latter is necessarily incomplete.  However in some cases these ties 
can be broken reliably.  For example, if the same score is returned for two 
dialects that only differ in the quote character, then we can check whether or 
not the quote character has any effect on the parsing result. If it doesn't, 
then the quote character only occurs \emph{inside} cells and does not affect 
the parsing outcome $\bft{C} = f_{\bfs{\theta}}^{-1}(\bft{x})$. The correct 
solution is then to ignore the quote character.  Similar tie breaking rules 
can be constructed for ties in the delimiter or the escape character.

\subsection{Potential Dialects}

In (\ref{eq:problem}) we select the best dialect from a set of dialects 
$\Theta_{\bft{x}}$. It's worthwhile to expand briefly on how 
$\Theta_{\bft{x}}$ is constructed. While in general it is the product set of 
all potential delimiters, quote characters, and escape characters in the file, 
there are small optimizations that can be done to shrink this parameter space.  
Doing this speeds up dialect detection and reduces erroneous results.

\begin{algorithm}[tb]
	\footnotesize
	\caption{Construction of potential dialects \label{alg:potential_dialects}}
	\begin{algorithmic}[1]
		\Function{GetDialects}{$\bft{x}$}
		\State $\bft{\tilde{x}} \gets$ \Call{FilterURLs}{$\bft{x}$}
		\State $\mathcal{D} \gets$ 
		\Call{GetDelimiters}{$\bft{\tilde{x}}$}
		\State $\mathcal{Q} \gets$ 
		\Call{GetQuotechars}{$\bft{\tilde{x}}$}
		\State $\mathcal{E}_{d,q} \gets \{\varepsilon\} \quad \forall 
		\theta_d, \theta_q \in \mathcal{D}\times\mathcal{Q}$
		\For {$i = 1, \ldots, \abs{\bft{\tilde{x}}} - 1$}
			\State $u, v \gets \bft{\tilde{x}}[i], \bft{\tilde{x}}[i+1]$
			\Comment{$\bft{\tilde{x}}[i]$ denotes the character in 
				$\bft{\tilde{x}}$ at position $i$.}
			\For {$\theta_d, \theta_q \in \mathcal{D} \times \mathcal{Q}$}
				\If {\Call{IsPotentialEscape}{u} {\bfseries and} $v \in \{\theta_d, 
					\theta_q\}$}
					\State $\mathcal{E}_{d,q} \gets \mathcal{E}_{d,q} \cup \{u\}$
				\EndIf
			\EndFor
		\EndFor
		\State $\Theta_{\bft{x}} \gets \emptyset$
		\For {$\theta_d, \theta_q \in \mathcal{D} \times \mathcal{Q}$}
			\For {$\theta_e \in \mathcal{E}_{d,q}$}
				\If {\textbf{not} \Call{MaskedByQuote}{$x, 
						\theta_d, \theta_q, 
						\theta_e$}}
					\State $\Theta_{\bft{x}} \gets 
					\Theta_{\bft{x}} \cup \{(\theta_d, 
					\theta_q, \theta_e)\}$
				\EndIf
			\EndFor
		\EndFor
		\State \Return $\Theta_{\bft{x}}$
		\EndFunction
	\end{algorithmic}
\end{algorithm}

Algorithm~\ref{alg:potential_dialects} presents pseudocode for the 
construction of $\Theta_{\bft{x}}$. The functions used in the algorithm are 
described in more detail in Appendix~\ref{app:additional_algorithms}. The 
algorithm proceeds as follows. As a preprocessing step, URLs are filtered from 
the file to remove potential delimiters that only occur in URLs.  Next, 
potential delimiters are found with {\footnotesize\textsc{GetDelimiters}} by 
comparing the Unicode category of each character with a set of allowed 
categories and filtering out characters that are explicitly blocked from being 
delimiters.  Next, potential quote characters are selected by matching with a 
set of allowed quote characters.  Subsequently, the escape characters are 
those that occur at least once before a delimiter or quote character and that 
belong to the Unicode ``Punctuation, other'' category 
\mycitep{unicode2018unicode}.  Finally, $\Theta_{\bft{x}}$ is the product set 
of the potential delimiters, quote characters, and escape characters, with the 
exception that dialects with delimiters that never occur outside the quote 
characters in the dialect are dropped.  This latter step removes more false 
positives from the set of potential dialects.

\section{Experiments}
\label{sec:comparison}

In this section we present the results of an extensive comparison study 
performed to evaluate our proposed method and existing alternatives. Since 
variability in CSV files is quite high and the number of potential CSV issues 
is large, an extensive study is necessary to thoroughly evaluate the 
robustness of each method. Moreover, since different groups of users apply 
different formats, it is important to consider more than one source of CSV 
files.

Our method presented above was created using a \emph{development set} of CSV 
files from two different corpora. The experimental section presents results 
from a comparison on an independent \emph{test set} that was unknown to the 
authors during the development of the method. This split aims to avoid 
overfitting of our method and report the accuracy of our method accurately.

In an effort to make our work transparent and reproducible, we release the 
full code to replicate the experiments through an online 
repository.\footnote{See 
	\url{https://github.com/alan-turing-institute/CSV_Wrangling}.}
This will also enable other researchers to easily build on our work and allow 
them to use our implementations of alternative detection methods.

\subsection{Data}

Data was collected from two sources: the UK government open data portal 
(\url{data.gov.uk}) and GitHub (\url{github.com}). These represent different 
groups of users (government employees vs. programmers) and it is expected that 
we find differences in both the format and the type of content of the CSV 
files. Data was collected by web scraping in the period of May/June 2018, 
yielding tens of thousands of CSV files. From these corpora of CSV files we 
randomly sampled a development set (3776 files from UKdata and 4536 files from 
GitHub).  These were used to develop and fine-tune the consistency measure 
presented above, and in particular were used to fine-tune the type detection 
engine.  

When development of the method was completed, an independent test set was 
sampled from the two sources (5000 files from each corpus). This test set is 
similar to the development data, with one exception.  During development we 
noticed that the GitHub corpus often contained multiple files from the same 
code repository.  These files usually have the same structure and dialect, 
thus representing essentially a single example. Therefore, during construction 
of the test set a limit of one CSV file per GitHub repository was put in 
place.  Thus we expect that the test set has greater variability and 
difficulty than the development set. It is worth emphasizing that the test set 
was not used in any way during the development of the method.

\subsection{Ground Truth}

To evaluate the detection method, we needed to obtain ground truth for the 
dialects of the CSV files. This is done through both automated and manual 
ways. The automated method is based on very strict functional tests that allow 
only simple CSV files with elementary cell contents. For instance, in one test 
we require that a file has a constant number of cells per row, no missing 
values, no nested quotes, etc. These automatic tests are sufficient to 
accurately determine the dialect of about a third of the CSV files. For the 
remainder of the files manual labelling was employed using a terminal-based 
annotation tool. Files that could not reasonably be considered CSV files were 
removed from the test set (i.e.  HTML, XML, or JSON files, or simple text 
files without any tabular data). The same holds for files for which no 
objective ground truth could be established, such as files formatted similarly 
to the example in Figure~\ref{fig:csv_weird}(c). After filtering out these 
cases the test set contained %
	4873
 files from GitHub.com 
and %
	4969
 files from the UK government open data portal. 

\subsection{Alternatives}

Since the dialect detection problem has not received much consideration in the 
literature, there are only a few alternative methods to compare to. We briefly 
present them here.

\subsubsection{Python Sniffer}

Python's built-in CSV module contains a so-called ``Dialect Sniffer'' that 
aims to automatically detect the dialect of the file\footnote{The dialect 
	sniffer was developed by Clifford Wells for his Python-DSV package 
	\mycitep{wells2002dsv} and was incorporated into Python version 2.3.}.  
This method detects the delimiter, the quote character, whether or not double 
quotes are used, and whether or not whitespace after the delimiter can be 
skipped.  There are two methods used to detect these properties.  The first 
method is used when quote characters are present in the file and detects 
adjacent occurrence of a quote character and another character (the potential 
delimiter).  In this method the quote character and delimiter that occur most 
frequently are chosen. The second method is used when there are no quote 
characters in the file. In this case a frequency table is constructed that 
indicates how often a potential delimiter occurs and in how many rows (i.e.  
comma occurred $x$ times in $y$ rows). The character that most often matches 
the expected frequency is considered the delimiter, and a fallback list of 
preferred delimiters is used when a tie occurs.  The detector also tries to 
detect whether or not double quoting is used within cells to escape a single 
quote character. This is done with a regular expression that can run into 
``catastrophic backtracking'' for CSV files that end in many empty delimited 
rows.  Therefore we place a timeout of two minutes on this detection method 
(normal operation never takes this long, so this restriction only captures 
this specific failure case).  As mentioned, this method tries to detect when 
whitespace following the delimiter can be stripped. We purposefully do not 
include this in our method as the CSV specification states, ``Spaces are 
considered part of a field and should not be ignored.'' 
\mycitep{shafranovich2005common}.

\subsubsection{HypoParsr}

HypoParsr \mycitep{dohmen2017multi} is the first dedicated CSV parser that 
takes the problem of dialect detection and messy CSV files into 
account.\footnote{%
	An R package for HypoParsr exists, but it was retracted from the R 
	package repository on the request of the package maintainer. We 
	nonetheless include the method in our experiments using the last 
	available version.} The method uses a hierarchy of possible parser 
configurations and a set of heuristics to try to determine which configuration 
gives the best result.  Unfortunately it is not possible use the HypoParsr R 
package to detect the dialect without running the full search that also 
includes header, table, and data type detection.  Therefore, we run the 
complete program and extract the dialect from the outcome. This means however 
that both the running time and any potential failure of the method are 
affected by subsequent parsing steps and not just by the dialect detection.  
This needs to be kept in mind when reviewing the results.  As the method can 
be quite slow, we add a timeout of 10 minutes per file.  Finally, the quote 
character in the dialect is not always reported faithfully in the final 
parsing result, since the underlying parser can strip quote characters 
automatically. We developed our own method to check what quote character was 
actually used during parsing.

\subsubsection{Wrangler}

In \mycitet{guo2011proactive} a table suitability metric is presented that 
balances consistency of cell types against the number of empty cells and cells 
with potential delimiters in them. This can therefore be used to detect the 
dialect of CSV files by selecting the dialect that does best on this metric.  
The suitability metric uses the concept of column type homogeneity, i.e. the 
sum of squares of the proportions of each data type in a column. Since the 
exact type detection method used in the paper is not available, we use our 
type detection method instead.

\subsubsection{Variations}

In addition to our complete data consistency measure, we also consider several 
variations to investigate the effect of each component. Thus, we include a 
method that only uses the pattern score and one that only uses the type score.  
We also include a variation that does not use tie-breaking.

\subsection{Evaluation}

The methods are evaluated on the accuracy of the full dialect as well as on 
the accuracy of each component of the dialect. Note that a method can either 
fail by detecting the dialect incorrectly or it can fail by not returning a 
result at all. The latter case can happen due to a timeout or an exception in 
the code (for the Python Sniffer or HypoParsr), or due to a tie in the scoring 
measure (for the Wrangler suitability metric or our method). Both types of 
failure are considered to represent an incorrect dialect detection.

\subsection{Results}

We describe the results by focusing on dialect detection accuracy, robustness 
of the methods, accuracy on messy CSV files, and runtime. Unless explicitly 
stated otherwise, all results reflect performance on the test dataset.

\subsubsection{Detection Accuracy}

\begin{table}[tb]
	\centering
	\begin{tabular}{@{}c@{}}
		\small
		\begin{tabular}{lrrrrrrrr}
Property & HypoParsr & Sniffer & Suitability & Pattern & Type & No Tie & Full\\
\hline
Delimiter & 87.32 & 86.95 & 65.57 & 92.67 & 88.02 & 91.50 & \textbf{94.91}\\
Quotechar & 83.03 & 92.47 & 45.02 & 95.38 & 90.21 & 93.95 & \textbf{97.35}\\
Escapechar & 87.83 & 94.34 & 74.78 & 97.95 & 96.24 & 95.57 & \textbf{99.22}\\
Overall & 80.61 & 85.66 & 38.60 & 91.16 & 83.38 & 90.72 & \textbf{93.76}\\
\hline
\end{tabular} \\[\abovecaptionskip]\\
		{\small \textbf{(a)} GitHub corpus}
	\end{tabular}
	\vskip\baselineskip
	\begin{tabular}{@{}c@{}}
		\small
		\begin{tabular}{lrrrrrrrr}
Property & HypoParsr & Sniffer & Suitability & Pattern & Type & No Tie & Full\\
\hline
Delimiter & 97.97 & 91.89 & 80.20 & 99.70 & 93.80 & 99.26 & \textbf{99.82}\\
Quotechar & 90.56 & 92.21 & 26.34 & 99.46 & 89.56 & 99.13 & \textbf{99.70}\\
Escapechar & 98.05 & 98.79 & 82.61 & \textbf{100.00} & 97.67 & 99.42 & 99.98\\
Overall & 90.44 & 90.84 & 25.32 & 99.40 & 87.18 & 99.11 & \textbf{99.68}\\
\hline
\end{tabular} \\[\abovecaptionskip]\\
		{\small \textbf{(b)} UKdata corpus}
	\end{tabular}
	\caption{Accuracy (in \%) of dialect detection for different methods 
		on both corpora. Complete failure of a detection method is 
		interpreted as an incorrect detection. ``Pattern'' and 
		``Type'' respectively indicate detection using only the 
		pattern score or only the type score. ``No Tie'' indicates our 
		method without tie-breaking. \label{tab:accuracies}}
\end{table}

The accuracy of dialect detection is shown in Tables~\ref{tab:accuracies}(a) 
and \ref{tab:accuracies}(b) respectively for the GitHub and UKdata corpora. We 
see that for both corpora and for all properties, our full data consistency 
method outperforms all alternatives. One exception occurs for the UKdata 
corpus, where the pattern-only score function yields a marginally higher 
accuracy on detecting the escape character. It is furthermore apparent from 
these tables that the GitHub corpus of CSV files is more difficult than the 
UKdata corpus.  This is also reflected in the number of dialects observed in 
these corpora: %
	8
 different dialects were found in 
the UKdata corpus vs. %
	35
 in the GitHub corpus. We 
postulate that this difference is due to the nature of the creators of these 
files. CSV files from the UK government open data portal are often created 
using standard tools such as Microsoft Excel or LibreOffice, and therefore are 
more likely to adhere to the CSV format \mycitep{shafranovich2005common}. On 
the other hand, the creators of the files in the GitHub corpus are more likely 
to be programmers and data scientists who may use non-standard or custom-made 
tools for importing and exporting their CSV files and may therefore use 
different formatting conventions.  It is interesting to note that even though 
the files in the UKdata corpus can be considered more ``regular'', we still 
achieve a considerable increase in detection accuracy over standard 
approaches.

Regarding the different variants of our method, we observe that the pattern 
score is in many cases almost as good as the full consistency measure.  This 
confirms our earlier statement that the pattern score is the main driver of 
the method and that the type score serves mainly to further improve the 
accuracy.  It is also clear that the type score alone does not suffice to 
accurately detect the dialect. The variant of our method that does not use 
tie-breaking yields a lower overall accuracy on both corpora, indicating the 
importance of tie-breaking in our method.  Additional result tables are 
presented in Appendix~\ref{app:results} that show the accuracy separated by 
human vs.  automatic detection of ground truth.  Unsurprisingly, all methods 
perform better on files for which the dialect could be detected through 
automatic means than on those that required human annotation.

\subsubsection{Failure}

\begin{figure}[tb]
	\centering
	\includegraphics[width=\textwidth]{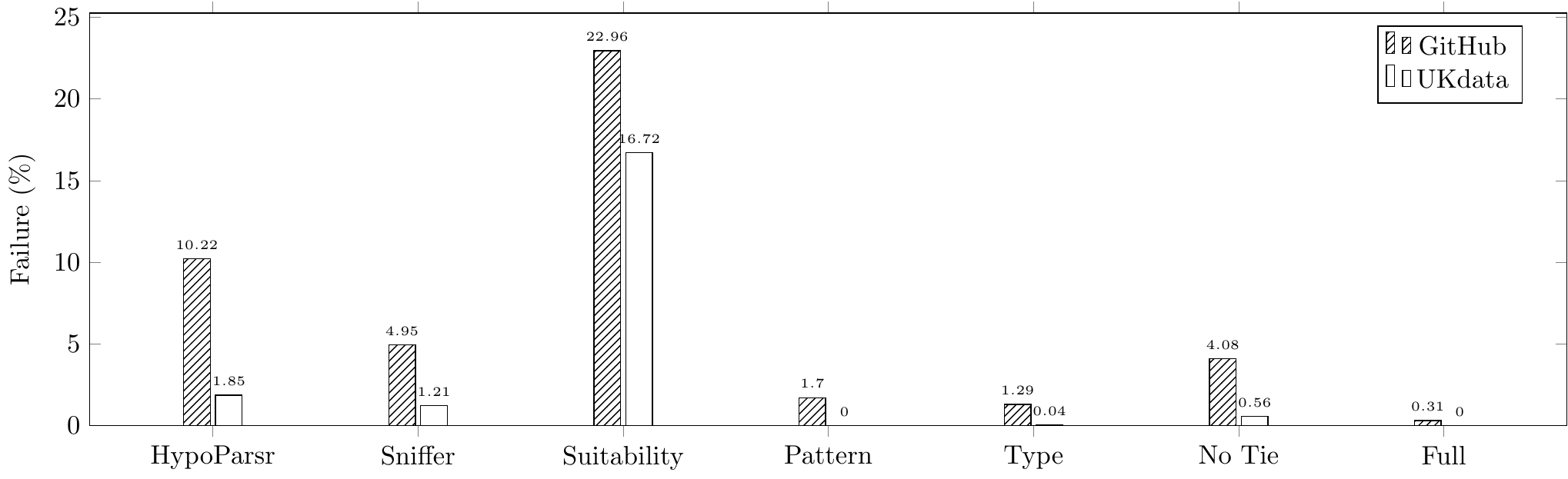}
	\caption{Percentage of failure for each of the detection methods on 
		both corpora. Note that this shows how often the detection 
		methods completely failed, not when it detected the parameters 
		incorrectly. \label{fig:failures}}
\end{figure}

Analysing the failure cases for the detection methods provides insight into 
their robustness. Figure~\ref{fig:failures} shows the proportion of files 
where no detection was made either due to a failure of the program or due to 
unbreakable ties in the scoring metric. This figure shows that HypoParsr 
\mycitep{dohmen2017multi} fails significantly more often on files from the 
GitHub corpus than files from the UKdata corpus.  This can be due to the fact 
that HypoParsr was developed using files from the UKdata corpus, and because 
these files are generally easier to parse. However, since it is not possible 
to only use HypoParsr for dialect detection, it could also be the case that 
failures occur in a later stage of the parsing process. In fact, investigating 
the failure cases for this method further we observe that no result was 
obtained in %
	62.7\%
 of the failures and that the 
timeout needed to be applied in %
	36.8\%
. Similarly, 
for the Python Sniffer these values are %
	77.5\%
 
and %
	22.5\%
 respectively. Note further that our 
proposed method fails on no files in the UKdata corpus and on only 
	0.31\%
 of files in the GitHub corpus. Failure in 
our method is only possible when ties occur, and the addition of both the type 
score and tie-breaking procedure ensures that this is rare. The high failure 
rate for the suitability metric from \mycitet{guo2011proactive} is exclusively 
due to ties in the scoring metric. Tie-breaking was not used here, as this is 
a feature of our method.


As Table~\ref{tab:accuracies}(a) shows, an incorrect detection in our method 
occurs mostly due to an incorrect detection of the delimiter. Analysing these 
failure cases in more detail reveals that of the files where the delimiter was 
detected incorrectly, the majority of files had the comma as the delimiter. In 
these cases, an incorrect detection occurred mostly because our method 
predicted the space as the delimiter. Some of these files indeed had a regular 
pattern of whitespace in the cells and received a high type score as well, 
causing the confusion. Other files had the comma as true delimiter, but had 
only one column of data. In these cases the true comma delimiter could be 
deduced by the human annotator from a header or because certain cells that 
contained the comma were quoted, but this type of reasoning is not captured by 
the data consistency measure.  In other failure cases, the pattern score 
predicted the correct delimiter, but the type score gave a low value, 
resulting in a low value of the consistency measure.  Some of these failure 
cases can certainly be addressed by improving the type detection procedure.

\subsubsection{Messy CSV Files}

\begin{table}[tb]
	\centering
	\begin{tabular}{@{}c@{}}
		\small
		\begin{tabular}{lrrrrrrrr}
 & HypoParsr & Sniffer & Suitability & Pattern & Type & No Tie & Full\\
\hline
Standard (3890) & 85.84 & 91.13 & 44.45 & 93.29 & 86.22 & 93.55 & \textbf{95.84}\\
Messy (983) & 59.92 & 63.99 & 15.46 & 82.71 & 72.13 & 79.55 & \textbf{85.55}\\
Total (4873) & 80.61 & 85.66 & 38.60 & 91.16 & 83.38 & 90.72 & \textbf{93.76}\\
\hline
\end{tabular} \\[\abovecaptionskip]\\
		{\small \textbf{(a)} GitHub corpus}
	\end{tabular}
	\vskip\baselineskip
	\begin{tabular}{@{}c@{}}
		\small
		\begin{tabular}{lrrrrrrrr}
 & HypoParsr & Sniffer & Suitability & Pattern & Type & No Tie & Full\\
\hline
Standard (4938) & 90.46 & 90.91 & 25.05 & 99.43 & 87.30 & 99.15 & \textbf{99.72}\\
Messy (31) & 87.10 & 80.65 & 67.74 & \textbf{93.55} & 67.74 & \textbf{93.55} & \textbf{93.55}\\
Total (4969) & 90.44 & 90.84 & 25.32 & 99.40 & 87.18 & 99.11 & \textbf{99.68}\\
\hline
\end{tabular} \\[\abovecaptionskip]\\
		{\small \textbf{(b)} UKdata corpus}
	\end{tabular}
	\caption{Accuracy (in \%) of dialect detection for different methods 
		on both corpora separated by standard and messy CSV files. The 
		numbers in parentheses represent the number of files in each 
		category. Complete failure of a detection method is again 
		interpreted as an incorrect detection. \label{tab:std_messy}}
\end{table}

Separating the files into those that are messy and those that follow the CSV 
standard further illustrates how our method improves over existing methods 
(see Table~\ref{tab:std_messy}). CSV files are considered ``standard'' when 
they use the comma as the delimiter, use either no quotes or the \verb+"+ 
character as quote character, and do not use the escape character. The table 
also highlights the difference in the amount of non-standard files in the 
different corpora and reiterates that the GitHub corpus contains more 
non-standard CSV files. This table also emphasises the significance of our 
contribution, with an increase of %
	21.3\%
 over 
the Python Sniffer for messy files averaged over both corpora. Note that on 
both corpora we also improve over current methods for \emph{standard} files.

\subsubsection{Runtime}

\begin{figure}[tb]
	\centering
	\includegraphics[width=\textwidth]{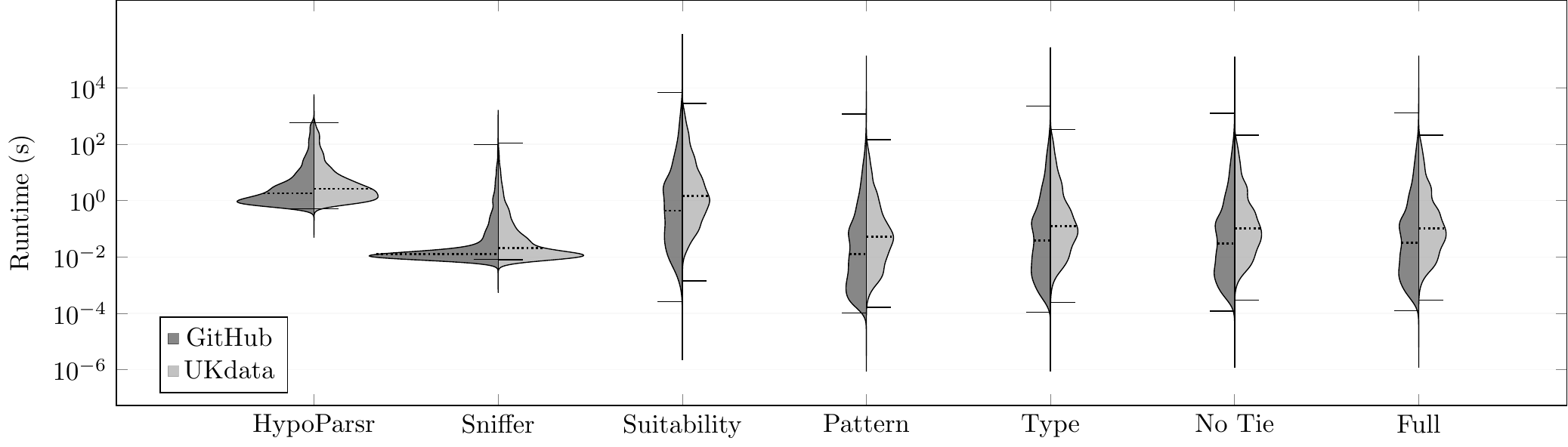}
	\caption{Runtime violin plots for both corpora. The whiskers show the 
		minimum and maximum values and the dashed lines show the 
		median.  See the note on HypoParsr in the main text.  
		\label{fig:runtimes}}
\end{figure}

Evaluating the runtime of each method is also relevant. 
Figure~\ref{fig:runtimes} shows violin plots for each method for both corpora.  
While the figures show that HypoParsr is the slowest detection method, this is 
not completely accurate. Since there is no way to use the method to only 
detect the dialect, the reported runtime is the time needed for the entire 
parsing process. The Python dialect sniffer is by far the fastest method and 
this can be most likely attributed to its simplicity in comparison to the 
other methods. Finally, all variations of our method have similar runtime 
characteristics and slightly outperform the Wrangler suitability metric. We 
note that our method has not been explicitly optimised for speed, and there 
are likely to be improvements that can be made in this aspect. Currently our 
method exhibits $O(|\bft{x}||\Theta_{\bft{x}}|)$ complexity.  Note however 
that the mean of the computation time for our method still lies well below one 
second, which is acceptable in almost all practical applications.

\section{Conclusion}%
\label{sec:conclusion}

We have presented a data consistency measure for detecting the dialect of CSV 
files. This consistency measure emphasizes a regular pattern of cells in the 
rows and favours dialects that yield identifiable data types. While we apply 
the consistency measure only to CSV dialect detection in this paper, it is 
conceivable that there are other applications of this measure outside this 
area. For instance, one can imagine identifying unstructured tables in HTML 
documents or in free text, or use the measure to locate the tables within CSV 
or Excel files.

The main challenge for today's data scientists is the inordinate amount of 
time spent on preparing data for analysis. One of the difficulties they face 
is importing data from messy CSV files that often require manual inspection 
and reformatting before the data can be loaded from the file. In this paper we 
have presented a method for automatic dialect detection of CSV files that 
achieves near-perfect accuracy on a large corpus of real-world examples, and 
especially improves the accuracy on messy CSV files.  This represents an 
important step towards automatically loading structured tabular data from 
messy sources.  As such, it allows many data scientists to spend less time on 
mundane data wrangling issues and more time on extracting valuable knowledge 
from their data.

\section*{Acknowledgements}

The authors would like to acknowledge the funding provided by the
UK Government's Defence \& Security Programme in support of the
Alan Turing Institute. The authors thank Chris Williams for useful 
discussions.

\bibliographystyle{plainnat}
\bibliography{references}

\appendix

\section{Data Type Detection}
\label{app:types}

As mentioned in the main text, we use a regular expression based type 
detection engine.  Below is a brief overview of the different types we 
consider and the detection method we use for that type. The order of the types 
corresponds to the order in which we evaluate the type tests, and we stop when 
a matching type is found.

\subsubsection*{Empty Strings}
Empty strings are considered a known type.

\subsubsection*{URLs and Email Addresses}
For this we use two separate regular expressions.

\subsubsection*{Numbers}
We consider two different regular expressions for numbers. First, we consider 
numbers that use ``digit grouping'', i.e. numbers that use a period or comma 
to separate groups of thousands.  In this case we allow numbers with a comma 
or period as thousands separator and allow for using a comma or period as 
radix point, respectively. Numbers in this form can not have E-notation, but 
can have a leading sign symbol. The second regular expression captures the 
numbers that do not use digit grouping.  These numbers can have a leading sign 
symbol (\verb-+- or \verb+-+), use a comma or period as radix point, and can 
use E-notation (i.e. \verb+123e10+).  The exponent in the E-notation can have 
a sign symbol as well.

\subsubsection*{Time}
Times are allowed in \verb+HH:MM:SS+, \verb+HH:MM+, and \verb+H:MM+ format.  
The AM/PM quantifiers are not included.

\subsubsection*{Percentage}
This corresponds to a number combined with the \verb+%+ symbol.

\subsubsection*{Currency}
A currency value is a number preceded by a symbol from the Unicode \verb+Sc+ 
category \mycitep{unicode2018unicode}.

\subsubsection*{Alphanumeric}
An alphanumeric string can follow two alternatives. The first alternative 
consists of first one or more number characters, then one or more letter 
characters, and then zero or more numbers, letters, or special characters. An 
example of this is the string \verb+3 degrees+. The second alternative first 
has one or more letter characters and then allows for zero or more numbers, 
letters, or special characters. An example of this is the string
\verb+NW1 2DB+. In both alternatives the allowed special characters are space, 
period, exclamation and question mark, and parentheses, including their 
international variants.

\subsubsection*{N/A}
While \verb+nan+ or \verb+NaN+ are caught in the alphanumeric test, we add 
here a separate test that considers \verb+n/a+ and \verb+N/A+.

\subsubsection*{Dates}
Dates are strings that are not numbers and that belong to one of forty 
different date formats. These date formats allow for the formats 
\verb+(YY)YYx(M)Mx(D)D+, \verb+(D)Dx(M)Mx(YY)YY+,\\ \verb+(M)Mx(D)Dx(YY)YY+ 
where \verb+x+ is a separator (dash, period, or space) and parts within 
parentheses can optionally be omitted. Additionally, the Chinese/Japanese date 
format\\ \begin{CJK}{UTF8}{gbsn}\verb+(YY)YY年(M)M月(D)D日+\end{CJK} and the 
Korean date format \begin{CJK}{UTF8}{mj}\verb+(YY)YY년(M)M월(D)D일+\end{CJK} 
are included.

\subsubsection*{Combined date and time}
These are formats for joint date and time descriptions. For these formats we 
consider\\ \verb+<date> <time>+ and \verb+<date>T<time>+ as well as those with 
a time zone offset appended.

\section{Additional Algorithm Descriptions}
\label{app:additional_algorithms}

Below we explicitly provide algorithms for functions mentioned in the main 
text that were omitted due to space constraints.

\subsection{Selecting Potential Delimiters}

Algorithm~\ref{alg:potential_delims} gives the function for selecting 
potential delimiters of the file $\bft{x}$. The function {\small 
	\textsc{FilterURLs}} replaces URLs in the file with a single letter, 
to avoid URLs generating spurious potential delimiters.  We use $\mathcal{B} = 
\{\verb+.+, \verb+/+, \verb+"+, \textrm{\textquotesingle}\}$ and
$\mathcal{C} = \{\verb+Lu+, \verb+Ll+, \verb+Lt+, \verb+Lm+, \verb+Lo+, 
\verb+Nd+, \verb+Nl+, \verb+No+, \verb+Ps+, \verb+Pe+, \verb+Cc+, \verb+Co+ 
\}$.

\begin{algorithm}[!ht]
	\footnotesize
	\caption{Selecting potential delimiters \label{alg:potential_delims}}
	\begin{algorithmic}[1]
		\Function{GetDelimiters}{$\bft{x}$, $\mathcal{C}$, 
			$\mathcal{B}$}
		\State $\bft{\tilde{x}} \gets$ \Call{FilterURLs}{$\bft{x}$}
		\State $\mathcal{D} \gets \{\varepsilon \}$
		\For {$x \in $ \Call{Unique}{$\bft{\tilde{x}}$}}
			\State $c \gets $ \Call{UnicodeCategory}{$x$}
			\If {$x = \texttt{Tab}$ \textbf{or} $(x \notin 
				\mathcal{B}$ \textbf{and} $c \notin 
				\mathcal{C})$}
				\State $\mathcal{D} \gets \mathcal{D} \cup 
				\{x\}$
			\EndIf
		\EndFor
		\State \Return $\mathcal{D}$
		\EndFunction
	\end{algorithmic}
\end{algorithm}
\FloatBarrier

\subsection{Selecting Potential Quote Characters}

The function {\small \textsc{FilterURLs}} is as described in the previous 
section.

\begin{algorithm}[!ht]
	\footnotesize
	\caption{Selecting potential quote characters}
	\begin{algorithmic}[1]
		\Function{GetQuotechars}{$\bft{x}$, $\mathcal{C}$, 
			$\mathcal{B}$}
		\State $\bft{\tilde{x}} \gets$ \Call{FilterURLs}{$\bft{x}$}
		\State $\mathcal{Q} \gets \{\textrm{\textquotesingle}, 
		\verb+"+, \verb+~+\} \cap \bft{\tilde{x}}$
		\State $\mathcal{Q} \gets \mathcal{Q} \cup \{\varepsilon \}$
		\State \Return $\mathcal{Q}$
		\EndFunction
	\end{algorithmic}
\end{algorithm}
\FloatBarrier

\subsection{Selecting Potential Escape Characters}

Potential escape characters are those in the Unicode \verb+Po+ (punctuation, 
other) category that are not explicitly blocked from being considered. For the 
latter we use a set\\ $\mathcal{W} = \{ \verb+!+, \verb+?+, \verb+"+, 
\textrm{\textquotesingle}, \verb+.+, \verb+,+, \verb+;+, \verb+:+, \verb+\%+, 
\verb+*+, \verb+&+, \verb+#+ \}$.

\begin{algorithm}[!ht]
	\footnotesize
	\caption{Selecting potential escape characters}
	\begin{algorithmic}[1]
		\Function{IsPotentialEscape}{$x$}
			\State $c \gets $ \Call{UnicodeCategory}{$x$}
			\If {$x \notin \mathcal{W}$ \textbf{and} $c = 
				\texttt{Po}$}
			\EndIf
			\State \Return \texttt{false}
		\EndFunction
	\end{algorithmic}
\end{algorithm}

\subsection{Masked by Quote Character}

The function {\small \textsc{MaskedByQuote}} in 
Algorithm~\ref{alg:potential_dialects} is used to prune the list of potential 
dialects by removing those where the delimiter never occurs outside a quoted 
environment. This function is straightforward to implement: it iterates over 
all characters in the file and keeps track of where a quoted section starts 
and ends, while taking quote escaping into account. Given this it is 
straightforward to check whether the given delimiter always occurs inside a 
quoted section or not.

\subsection{Parser}
\label{app:parser}

The code we use for our CSV parser $f^{-1}_{\bfs{\theta}}$ borrows heavily 
from the CSV parser in the Python standard library, but differs in a few small 
but significant ways.  First, our parser only interprets the escape character 
if it proceeds the delimiter, quote character, or itself. In any other case 
the escape character serves no purpose and is treated as any other character 
and is not dropped.  Second, our parser only strips quotes from cells if they 
surround the entire cell, not if they occur within cells. This makes the 
parser more robust against misspecified quote characters. Finally, when we are 
in a quoted cell we automatically detect double quoting by looking ahead 
whenever we detect a quote, and checking if the next character is \emph{also} 
a quote character.  This enables us to drop double quoting from our dialect 
and only marginally affects the complexity of the code.

\FloatBarrier

\section{Additional Results}
\label{app:results}

\begin{table}[ht]
	\centering
	\begin{tabular}{@{}c@{}}
		\footnotesize
		\begin{tabular}{lrrrrrrrr}
Property & HypoParsr & Sniffer & Suitability & Pattern & Type & No Tie & Full\\
\hline
Delimiter & 83.09 & 82.15 & 59.94 & 89.71 & 84.41 & 88.26 & \textbf{93.15}\\
Quotechar & 76.41 & 89.50 & 39.99 & 93.49 & 85.89 & 91.16 & \textbf{96.05}\\
Escapechar & 83.87 & 92.44 & 73.84 & 97.10 & 94.80 & 93.45 & \textbf{98.72}\\
Overall & 72.56 & 80.09 & 29.46 & 87.51 & 76.92 & 87.04 & \textbf{91.33}\\
\hline
\end{tabular} \\[\abovecaptionskip]\\
		{(a) GitHub -- human annotated}
	\end{tabular}
	\vskip\baselineskip
	\begin{tabular}{@{}c@{}}
		\footnotesize
		\begin{tabular}{lrrrrrrrr}
Property & HypoParsr & Sniffer & Suitability & Pattern & Type & No Tie & Full\\
\hline
Delimiter & 93.87 & 94.40 & 74.29 & 97.28 & 93.61 & 96.54 & \textbf{97.64}\\
Quotechar & 93.30 & 97.07 & 52.83 & 98.32 & 96.91 & 98.27 & \textbf{99.37}\\
Escapechar & 93.98 & 97.28 & 76.23 & 99.27 & 98.48 & 98.85 & \textbf{100.00}\\
Overall & 93.09 & 94.29 & 52.77 & 96.81 & 93.40 & 96.44 & \textbf{97.54}\\
\hline
\end{tabular} \\[\abovecaptionskip]\\
		{(b) GitHub -- automatically annotated}
	\end{tabular}
	\vskip\baselineskip
	\begin{tabular}{@{}c@{}}
		\footnotesize
		\begin{tabular}{lrrrrrrrr}
Property & HypoParsr & Sniffer & Suitability & Pattern & Type & No Tie & Full\\
\hline
Delimiter & 97.31 & 92.17 & 81.91 & 99.59 & 93.16 & 98.95 & \textbf{99.74}\\
Quotechar & 87.46 & 92.93 & 24.61 & 99.30 & 87.49 & 98.77 & \textbf{99.56}\\
Escapechar & 97.43 & 99.65 & 85.21 & \textbf{100.00} & 96.76 & 99.18 & 99.97\\
Overall & 87.28 & 91.14 & 23.12 & 99.21 & 84.71 & 98.74 & \textbf{99.53}\\
\hline
\end{tabular}\\[\abovecaptionskip]\\

		{(c) UKdata -- human annotated}
	\end{tabular}
	\vskip\baselineskip
	\begin{tabular}{@{}c@{}}
		\footnotesize
		\begin{tabular}{lrrrrrrrr}
Property & HypoParsr & Sniffer & Suitability & Pattern & Type & No Tie & Full\\
\hline
Delimiter & 99.42 & 91.28 & 76.42 & 99.94 & 95.22 & 99.94 & \textbf{100.00}\\
Quotechar & 97.42 & 90.63 & 30.17 & 99.81 & 94.12 & 99.94 & \textbf{100.00}\\
Escapechar & 99.42 & 96.90 & 76.87 & \textbf{100.00} & 99.68 & 99.94 & \textbf{100.00}\\
Overall & 97.42 & 90.18 & 30.17 & 99.81 & 92.64 & 99.94 & \textbf{100.00}\\
\hline
\end{tabular}\\[\abovecaptionskip]\\

		{(d) UKdata -- automatically annotated}
	\end{tabular}
	\caption{Accuracy (in \%) of dialect detection for different methods 
		on both corpora, split between files that were manually 
		annotated and files that where the dialect was determined 
		through automatic means. As in the main text, failure of a 
		method to return a result is interpreted as an incorrect 
		detection.  \label{tab:additional_accuracy}}
\end{table}

\end{document}